\documentclass{article}
\usepackage{graphicx}
\def\no{\noindent}
\def\bc{\begin{center}}
\def\ec{\end{center}}

\def\beq{\begin{equation}}
\def\eeq{\end{equation}}

\def\sm{\sigma^{\min}}

\title{On the minimal conductivity of graphene}

\author{K. Ziegler\\
Institut f\"ur Physik, Universit\"at Augsburg\\
D-86135 Augsburg, Germany}

\begin{document}

\maketitle

\no
Abstract:

\no
The minimal conductivity of graphene is a quantity measured in the DC limit.
It is shown, using the Kubo formula, that the actual value of the minimal conductivity 
is sensitive to the order in which certain limits are taken.
If the DC limit is taken before the integration over energies 
is performed, the minimal conductivity of graphene is $4/\pi$ (in units of $e^2/h$) and 
it is  $\pi/2$ in the reverse order. The value $\pi$ is obtained
if weak disorder is included via a small frequency-dependent selfenergy.
In the high-frequency limit the minimal conductivity approaches $\pi/2$ and drops to zero
if the frequency exceeds the cut-off energy of the particles.    

\vskip1cm

{\it Introduction}.
The conductivity $\sigma_{\mu\mu}$ of graphene varies with the density of quasiparticles almost linearly 
with a minimal value $\sm\approx 4 e^2/h$ \cite{novoselov05,zhang05}. 
In terms of theoretical calculations, there has been some confusion about the actual value of 
the minimal conductivity $\sm$. 
This confusion is twofold: one originates from the experimentally 
observed value that is roughly three times as big as most of the calculated values, the other one 
is related to the theoretical calculations that produced different
values of $\sm$ (calculated per spin and per valley):
\beq
\sm_1={1\over\pi}{e^2\over h} 
\hskip0.5cm
\cite{ludwig94,ziegler98,katsnelson05,gusynin05,peres06,tworzydlo06,cserti06,ostrowsky06,ryu06},
\hskip0.5cm
\sm_2={\pi\over 8} {e^2\over h}
\hskip0.5cm
\cite{ludwig94,cserti06,falkovsky06},
\hskip0.5cm
\sm_3={\pi\over 4} {e^2\over h}
\hskip0.5cm
\cite{ziegler06} .
\label{results}
\eeq
$\sm_1$ was obtained from the Kubo formula \cite{ludwig94,ziegler98,gusynin05,peres06,ostrowsky06,ryu06}
as well as from the Landauer formula \cite{katsnelson05,tworzydlo06,ryu06}, whereas $\sm_{2,3}$ were
obtained from the Kubo formula only. All these results were calculated near the ballistic regime
of the quasiparticles. The possibility to reach the experimentally observed values of the
minimal conductivity by including long-range disorder due to charged impurities was also 
discussed recently \cite{nomura06}. The latter will not be considered in the subsequent discussion.
Instead, it shall be explained that all the results in Eq. (\ref{results}) can be obtained from 
the standard Kubo formula of nearly ballistic quasiparticles by taking limits in different order.
When a non-zero temperature $T$ is considered, the conductivity is a function $\sm(\omega/T,\eta/T)$, 
for frequency $\omega$ and scattering rate $\eta$.


{\it Frequency-dependent conductivity}.
The conductivity $\sigma_{\mu\nu}$ is given by the Kubo formula as a response to an
external field with frequency $\omega$.
Here the representation given in Ref. \cite{ziegler06} is used
\beq
\sigma_{\mu\nu}=i{e^2\over\hbar}
\int\int Tr\Big\{[H,r_\mu]
\delta(H-\epsilon')[H,r_\nu]\delta(H-\epsilon)
\Big\}
{1\over \epsilon-\epsilon'+\omega-i\alpha}
{f_\beta(\epsilon')-f_\beta(\epsilon)
\over\epsilon-\epsilon'}d\epsilon d\epsilon' ,
\label{cond00}
\eeq
where $f_\beta(\epsilon)=1/(1+\exp(\beta\epsilon))$ is the Fermi function
at temperature $T=1/(k_B\beta)$.
For the minimal conductivity only the real part of the diagonal 
conductivity $\sigma_{\nu\nu}'=Re(\sigma_{\nu\nu})$ is of interest. 
After taking the limit $\alpha\to0$,
the $\epsilon'$ integration can be performed and gives
\beq
\sigma_{\nu\nu}'=\pi {e^2\over\hbar}
\int Tr\Big\{[H,r_\nu]
\delta(H-\epsilon-\omega)[H,r_\nu]\delta(H-\epsilon)
\Big\}
{f_\beta(\epsilon+\omega)-f_\beta(\epsilon)
\over\omega}d\epsilon .
\label{tcond}
\eeq
In the zero-temperature limit $\beta\to\infty$ this becomes 
\beq
\sigma_{\nu\nu}'=-\pi {e^2\over\hbar}{1\over\omega}
\int_{-\omega/2}^{\omega/2} Tr\Big\{[H,r_\nu]
\delta(H-\omega/2-\epsilon)[H,r_\nu]\delta(H+\omega/2-\epsilon)
\Big\}
d\epsilon .
\label{cond3}
\eeq
The minimal conductivity $\sm$ is obtained by taking the limit $\omega\to0$
of $\sigma_{\nu\nu}'$. It is tempting to ignore the $\omega$ dependence of the integrand
and replace the right-hand side by the integrand at $\epsilon=\omega=0$.
It will be shown subsequently that this does not agree with the result, when we perform
the energy integration first and take the limit $\omega\to0$ later.


{\it Dirac fermions}.
The Hamiltonian of Dirac fermions in 2D with wavevector $(k_1,k_2)$
\beq
H=\sigma_1k_1+\sigma_2k_2 
\label{ham1}
\eeq
describes the low-energy quasiparticles in graphene.
$\sigma_j$ ($j=1,2,3$) are Pauli matrices. $H$ can be diagonalized 
as $diag(k,-k)$ with $k=\sqrt{k^2_1+k^2_2}$.
The current operator transforms under Fourier transformation as
\[
j_\mu=-ie[H,r_\mu]\longrightarrow e{\partial H\over\partial k_\mu}  .
\]
This means that the current operators for the Hamiltonian $H$ in 
Eq. (\ref{ham1}) is a $2\times2$ matrix with vanishing diagonal elements.
The representation of $j_2$ in terms of energy eigenstates reads
\beq
j_2={e\over k}\pmatrix{
k_2 & ik_1 \cr
-ik_1 & -k_2\cr
}
.
\label{current2}
\eeq
Thus, the current $j_2$ does not depend on $k$ but only on the polar angle.
The trace term in the conductivity of Eq. (\ref{cond3}) reads 
\[
T(\epsilon)=-Tr\Big\{[H,r_\mu]
\delta(H-\omega/2-\epsilon)[H,r_\mu]\delta(H+\omega/2-\epsilon)
\Big\}
\]
\beq
=\int Tr_2\Big\{
{\partial H\over\partial k_\mu}\delta(H-\omega/2-\epsilon)
{\partial H\over\partial k_\mu}\delta(H+\omega/2-\epsilon)
\Big\}{d^2k\over (2\pi)^2} ,
\label{t1}
\eeq
where $Tr_2$ is the trace with respect to $2\times2$ matrices.
After diagonalizing $H$ this becomes together with the current in Eq. (\ref{current2})
\[
T(\epsilon)=\int {k_1^2\over k^2}
[\delta(k+\omega/2+\epsilon)\delta(k+\omega/2-\epsilon)
+\delta(k-\omega/2-\epsilon)\delta(k-\omega/2+\epsilon)]
{dk_1dk_2\over (2\pi)^2}
\]
\beq
+\int {k_2^2\over k^2} 
[\delta(k-\omega/2-\epsilon)\delta(k+\omega/2-\epsilon)
+\delta(k+\omega/2+\epsilon)\delta(k-\omega/2+\epsilon)]
{dk_1dk_2\over(2\pi)^2} 
\label{trace1}
\eeq
which is a symmetric function with respect to $\epsilon$.

Now a soft Dirac delta function $\delta_\eta(x)$ is considered with
\beq
\delta_\eta(x)={1\over\pi}{\eta\over x^2+\eta^2}=
-{1\over 2i\pi}\left[
{1\over x+i\eta}-{1\over x-i\eta}
\right] .
\label{deltas}
\eeq
The parameter $\eta$ (a scattering rate) can be understood as the imaginary part
of the selfenergy, created, for instance, by random fluctuations due to disorder
\cite{ziegler98}. 
With the energy cut-off $\lambda$ for the Dirac fermions and $\eta\sim0$, the integral of the
double product of soft Dirac delta functions reads
\[
\int_0^\lambda\delta_\eta(k-a)\delta_\eta(k-b)kdk
\sim (a+b)\delta_\eta (a-b){1\over 8}\left[
\Theta(\lambda-a)+\Theta(a)+\Theta(\lambda-b)+\Theta(b) -2
\right]
\]
\[
-{\eta\over a-b}{1\over 2\pi}\left[
\Theta(\lambda-b)+\Theta(b)-\Theta(\lambda-a)-\Theta(a)
\right] .
\]
Returning to Eq. (\ref{trace1}), we restrict the variable $\epsilon$
to $-\omega/2<\epsilon<\omega/2$, since for low temperature the term
\[
f_\beta(\epsilon+\omega/2)-f_\beta(\epsilon-\omega/2)
=-{\sinh(\beta\omega/2)\over \cosh(\beta\omega/2)+\cosh(\beta\epsilon)}
\]
in the conductivity is exponentially small for $|\epsilon|>\omega/2$. 
If it is further assumed that $\eta,\omega\ll\lambda$ we obtain
\beq
T(\epsilon)\sim{\pi\over(2\pi)^2}\left(
{\omega\over4}\delta_{\eta}(\epsilon)+{\eta\over\pi\omega}
\right)\Theta(\lambda-\omega/2) ,
\label{t2}
\eeq
where the prefactor $\pi$ is a result of the angular integration of
$k_j^2/k^2$.
The first term describes interband scattering (i.e. scattering between states
with different energies $\pm\epsilon$) and the second term intraband scattering
(i.e. scattering between states with the same energy $\epsilon$ or $-\epsilon$).
The intraband scattering term increases linearly with the scattering rate $\eta$,
in contrast to the $\eta$-independent interband scattering. The frequency dependence
is also different for the two types of scattering: the interband term increases
with $\omega$, whereas the intraband term decreases.

The temperature-dependent conductivity can be calculated from Eqs. (\ref{tcond})
and (\ref{t1}) as
\beq
\sigma_{22}'=-\pi {e^2\over\hbar}\int T(\epsilon)
{f_\beta(\epsilon+\omega/2)-f_\beta(\epsilon-\omega/2)
\over\omega}d\epsilon .
\eeq
Thus Eq. (\ref{t2}) implies
\beq
\sigma_{22}'\sim
-{\pi e^2\over 8h}[f_\beta(\omega/2)-f_\beta(-\omega/2)]
+{e^2\over 2h}{\eta\over\omega^2}
\int_{-\omega/2}^{\omega/2}{\sinh(\beta\omega/2)\over \cosh(\beta\epsilon)+\cosh(\beta\omega/2)}d\epsilon 
\label{cond2}
\eeq
for $\omega<2\lambda$ and a vanishing conductivity for $\omega>2\lambda$.
The integral in the second term gives
\[
{1\over\omega}\int_{-\omega/2}^{\omega/2}{\sinh(\beta\omega/2)\over 
\cosh(\beta\epsilon)+\cosh(\beta\omega/2)}d\epsilon
={4\over\beta\omega}{\rm arctanh}\left[
\tanh^2(\beta\omega/4)\right] .
\]
Moreover, the relation
\[
{\rm arctanh}(x)={1\over 2}\log\left[{1+x\over1-x}\right]
\]
can be used to get 
\beq
\sigma_{22}'\sim
{\pi e^2\over 8h}
\tanh(\beta\omega/4)
+{e^2\over h}
{\beta\eta\over(\beta\omega)^2}\log\left[{1+\tanh^2(\beta\omega/4)
\over 1-\tanh^2(\beta\omega/4)}
\right] .
\label{cond3a}
\eeq
This is the main result of this paper. It shows that 
the conductivity depends on two parameters, $\beta\omega$ and $\beta\eta$.
Experimentally interesting is the case where $\beta\eta$ is fixed and $\beta\omega$
is varied (cf. Fig. 1). This is motivated by two facts. The first one
is related to the origin of $\eta$ (the scattering rate or inverse scattering time).
It grows with increasing disorder. An important source of disorder in graphene are 
``ripples'' in the carbon sheet \cite{geim07,meyer07} which are created by thermal
fluctuations. Therefore, a simple estimate gives a linear growth of the scattering
rate with temperature. The other support for a
constant $\beta\eta$ with respect to temperature comes from the experimentally observed 
constant minimal conductivity, found for a wide range of temperatures 
\cite{novoselov05}. It will be shown below that the minimal conductivity at $\omega=0$
and $\beta<\infty$ is proportional to $\beta\eta$.
 
{\it Discussion of the results: I) zero temperature.}
With expression (\ref{t2}), the conductivity $\sigma_{22}'$ in Eq. (\ref{cond3}) 
reads eventually
\beq
\sigma_{22}' ={e^2\over4\hbar}{1\over\omega}
\int_{-\omega/2}^{\omega/2}\left(
{\omega\over4}\delta_{\eta}(\epsilon)+{\eta\over\pi\omega}
\right)d\epsilon
={\pi\over 8}\left(
1+{4\eta\over\pi\omega}
\right){e^2\over h} .
\label{condf}
\eeq
The linear increase with the scattering rate $\eta$ is in agreement with the Drude formula
\[
\sigma'={\sigma_0/\eta\over 1+(\omega/\eta)^2} 
\]
for large $\omega$, except for the different power in $\omega$.
This dependence on $\eta$ is in qualitative agreement with the reduction of the minimal conductivity 
after annealing (i.e. effectively reducing $\eta$) which was 
observed in a recent experiment by Geim and Novoselov \cite{geim07}. 

The expression of the conductivity $\sigma_{22}'$ can be studied in
several limits. As a first example, in Eq. (\ref{trace1}) the limit $\omega\to0$ 
is taken first and then the limit $\eta\to0$. Then the conductivity in 
Eq. (\ref{cond3}) reads
\beq
\sm_1 ={e^2\over\hbar}{\eta^2\over\pi^2}\int_0^\infty{1\over (k^2+\eta^2)^2}
kdk\approx{1\over\pi}{e^2\over h} .
\label{si1}
\eeq
In the next example the result in Eq. (\ref{condf}) is considered. 
This yields for $\eta\ll\omega$ the minimal conductivity
\beq
\sm_2\approx {\pi\over 8}{e^2\over h}
\hskip0.5cm 
{\rm for}\ \ \eta\approx0
\label{si2}
\eeq
and 
\beq
\sm_3\approx {\pi\over 4}{e^2\over h}
\hskip0.5cm 
{\rm for}\ \ \eta\approx\omega.
\label{si3}
\eeq
The last result agrees reasonably well with the experimental observation of Ref.
\cite{novoselov05} if it is multiplied by 4, the factor that is taking care of
the two-fold spin and the two-fold valley degeneracy of graphene.

{\it II) frequency and temperature dependence.}
There are two asymptotic regimes with
\[
{1\over(\beta\omega)^2}\log\left[{1+\tanh^2(\beta\omega/4)
\over 1-\tanh^2(\beta\omega/4)}
\right] \sim\cases{
1/8 & for $\beta\omega\sim0$ \cr
1/2\beta\omega 
& for $\beta\omega\sim\infty$ \cr
}
\]
which implies for the conductivity
\beq
\sigma_{22}'\sim {e^2\over 8h}\cases{
\beta\eta  & for $\beta\omega\sim0$ \cr
\pi +4\beta\eta/\beta\omega & for $\beta\omega\sim\infty$ \cr
} .
\label{asympt}
\eeq
The result of Eq. (\ref{condf}) is reproduced when 
the temperature is sent to zero first.
Experimentally, however, it is more realistic
to study the DC conductivity at a nonzero temperature. Then the conductivity
depends on the scattering rate as $\sigma_{22}'\propto\beta\eta$. 
Remarkable is the frequency-dependent conductivity in comparison with the 
Drude formula.
The Drude conductivity vanishes for large frequencies like $\propto \omega^{-2}$, in contrast to the
almost constant behavior in Eq. (\ref{asympt}) for $\omega<2\lambda$ and an abrupt vanishing of
the conductivity if the frequency exceeds the energy cut-off of the Dirac fermions $2\lambda$. 

\begin{figure}
\centering
\includegraphics[width=0.6\textwidth]{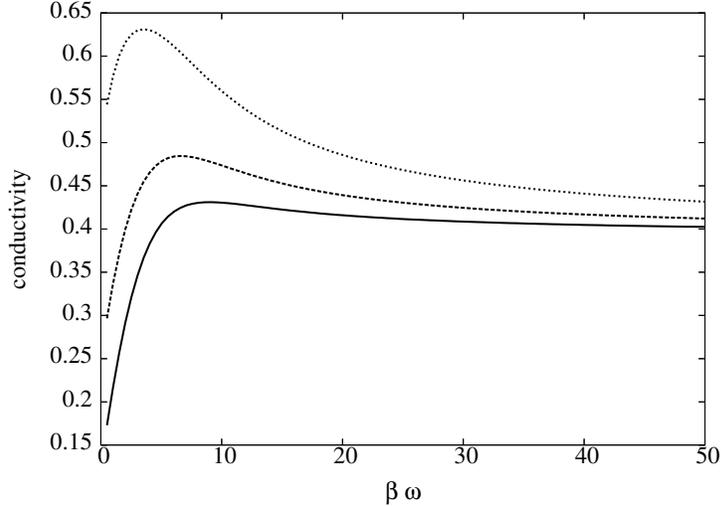}
\caption{Conductivity of Dirac fermions vs. $\beta\omega$ ($\beta$ is the inverse temperature and
$\omega$ the frequency) in units of $e^2/h$. The conductivity
increases with the rate $\beta\eta=1,2,4$ ($\eta$ the scattering rate).
It is assumed that $\beta\eta$ does not depend on $\beta$ (from Eq. (\ref{cond3a})).}
\label{plotconda}
\end{figure}

In conclusion, the Kubo formula produces a non-universal value for the minimal
conductivity in graphene. Depending on the order of the limits, this quantity can vary
over a wider range in units of $e^2/h$. The frequency dependent conductivity
of graphene at the Dirac point is exceptional, since it is almost constant and drops
to zero when the frequency reaches the energy cut-off of the Dirac fermions.

\vskip0.3cm
\no
I am grateful to S. Mikhailov for interesting discussions and Ch. Mudry for bringing
Ref. \cite{ryu06} to my attention.


\begin{thebibliography}{99}

\bibitem{novoselov05}
K.S. Novoselov, A.K. Geim, S.V. Morozov, D. Jiang, M.I. Katsnelson, I.V. Grigorieva,
and A.A. Firsov, Nature {\bf 438}, 197 (2005)

\bibitem{zhang05}
Y. Zhang, V.-W. Tan, H.L. St\"ormer, and P. Kim, Nature {\bf 438}, 201 (2005)

\bibitem{ludwig94}
A.W.W. Ludwig, M.P.A. Fisher, R. Shankar, and G. Grinstein, 
Phys. Rev. B {\bf 50}, 7526 (1994)

\bibitem{ziegler98}
K. Ziegler, Phys. Rev. B {\bf 55}, 10661 (1997);
Phys. Rev. Lett. {\bf 80}, 3113 (1998)

\bibitem{katsnelson05}
M.I. Katsnelson, 
Eur. Phys. J. B {\bf 51}, 157 (2006)

\bibitem{gusynin05}
V.P. Gusynin and S.G. Sharapov, Phys. Rev. B {\bf 66}, 045108 (2002);
Phys. Rev. Lett. {\bf 95}, 146801 (2005); 
Phys. Rev. B {\bf 73}, 245411 (2006)

\bibitem{peres06}
N.M.R. Peres, F. Guinea, and A.H. Castro Neto, Phys. Rev. 
B {\bf 73}, 125411 (2006)

\bibitem{tworzydlo06}
J. Tworzydlo, B. Trauzettel, M. Titov, A. Rycerz, and C. Beenakker, 
Phys. Rev. Lett. {\bf 96}, 246802 (2006)

\bibitem{cserti06}
J. Cserti, Phys. Rev. B {\bf 75}, 033405 (2007)

\bibitem{ostrowsky06}
P.M. Ostrovsky, I.V. Gornyi, A.D. Mirlin, 
Phys. Rev. B {\bf 74}, 235443 (2006)

\bibitem{ryu06}
S. Ryu, C. Mudry, A. Furusaki, A.W.W. Ludwig,
cond-mat/0610598

\bibitem{falkovsky06}
L.A. Falkovsky and A.A. Varlamov, cond-mat/0606800

\bibitem{ziegler06}
K. Ziegler, Phys. Rev. Lett. {\bf 97}, 266802 (2006)

\bibitem{nomura06}
K. Nomura and A.H. MacDonald, 
Phys. Rev. Lett. {\bf 96}, 256602 (2006)

\bibitem{geim07}
A.K. Geim and K.S. Novoselov, Nature Materials, {\bf 6}, 183 (2007)

\bibitem{meyer07}
J.C. Meyer, A. K. Geim, M. I. Katsnelson, K. S. Novoselov, T. J. Booth, 
and S. Roth, cond-mat/0701379

\end{thebibliography}
\end{document}